\documentclass[12pt]{article}

\usepackage{amsmath}
\usepackage{latexsym}
\newcommand\be{\begin{equation}}
\newcommand\ee{\end{equation}}
\newcommand\bea{\begin{eqnarray}}
\newcommand\eea{\end{eqnarray}}
\newcommand\beas{\begin{eqnarray*}}
\newcommand\eeas{\end{eqnarray*}}

\usepackage[unicode=true,pdfusetitle,
 bookmarks=true,bookmarksnumbered=false,bookmarksopen=false,
 breaklinks=false,pdfborder={0 0 1},backref=false,colorlinks=false]
 {hyperref}

\def\tr{{\rm Tr}}

\def\Xint#1{\mathchoice
{\XXint\displaystyle\textstyle{#1}}%
{\XXint\textstyle\scriptstyle{#1}}%
{\XXint\scriptstyle\scriptscriptstyle{#1}}%
{\XXint\scriptscriptstyle\scriptscriptstyle{#1}}%
\!\int}
\def\XXint#1#2#3{{\setbox0=\hbox{$#1{#2#3}{\int}$ }
\vcenter{\hbox{$#2#3$ }}\kern-.5\wd0}}

\def\dashint{\Xint-}

\begin{document}
\title{Large N Matrix Hyperspheres and the Gauge-Gravity Correspondence\footnote{WITS-CTP-147}}

\author{\\
Mthokozisi Masuku\footnote{Email: Mthokozisi.Masuku@students.wits.ac.za}, Mbavhalelo Mulokwe\footnote{Email: Mbavhalelo.Mulokwe@students.wits.ac.za}  and Jo\~ao P. Rodrigues\footnote{Email: Joao.Rodrigues@wits.ac.za} \\
\\
National Institute for Theoretical Physics \\
School of Physics and Centre for Theoretical Physics \\
University of the Witwatersrand, Johannesburg\\
Wits 2050, South Africa 
}

\maketitle

\begin{abstract}
The large $N$ dynamics of a subsector of $d=0$ interacting complex multi matrix systems, which is naturally parametrized by a matrix valued radial coordinate, and which embodies the canonical AdS/CFT relationship between 
't Hooft's coupling constant and radius, is obtained. Unlike the case of the single complex matrix, for two or more complex matrices a new repulsive logarithmic potential is present, and as a result the density of radial eigenvalues has suport on an hyper annulus.  For the single complex matrix, the integral over the angular degrees of freedom of the Yang-Mills interaction can be carried out exactly, and in the presence of an harmonic potential, the density of radial eigenvalues is shown to be of the Wigner type.
\end{abstract}

\newpage

\section{Introduction}
The AdS/CFT correspondence \cite{Maldacena:1997re}, \cite{Gubser:1998bc}, \cite{Witten:1998qj} has provided new insights into the properties of several different gravitational objects from their corresponding description in terms of  the large $N$ limit of of matrix valued (super) Yang-Mills theories. It is of great interest to understand how or if gravitational properties emerge from the large $N$ limit of matrix theories, including in settings without supersymmetry or even conformal invariance. Of particular interest is the emergence of a semiclassical / geometric background in the large $N$ limit of matrix theories at strong coupling. 

 Two defining relationships of the AdS/CFT correspondence in the canonical case are
 
 \bea
\label{AdSOne} R &\sim& \lambda^{\frac{1}{4}}    \\
\label{AdSTwo} g_{YM}^2&\sim& g_s
\eea

\noindent 
where $R$ is the radius of both the $AdS_5$ and $S^{5}$ of the $AdS_{5}\times S^{5}$  background, $g_{YM}$ is the (super) Yang-Mills coupling, $\lambda= g_{YM}^2 N$ the 't Hooft coupling and $g_s$ the Type IIB string coupling. 

In this communication, we consider several properties associated with the large $N$ limit of the integral

\be\label{FIntegral}
             Z = \int [dX_I] e^{-S} \ , S= \frac{w^2}{2 g_{YM}^2} \sum_{I=1}^6 \tr X_I^2 - \frac{1}{g_{YM}^2} \sum_{I\ne J}^6 \tr ([X_I,X_J][X_I,X_J]) 
\ee

\noindent
which corresponds to the leading compactification ( e.g., or $S^3 \times T^1$ or $S^4$) of the bosonic Higgs sector of $\cal{N}=$ $4$ SYM. The harmonic  "frequency"  $w$ is a function of the curvature of the manifold, and serves as an infrared regulator. $g_{YM}$ "has been scaled out of the action", so that the relationship (\ref{AdSTwo}) is apparent. 

For this system, the relationship (\ref{AdSOne}) now follows from dimensional analysis\footnote{We extend to $d=0$ the standard scalar dimensions $[X_I]=\frac{d-2}{2}$}, in that 

\be
[X_I]  \sim \lambda^{\frac{1}{4}} 
\ee

If a suitable matrix "radial coordinate" can be identified, it will satisfy  (\ref{AdSOne}) simply by dimensional analysis, provided no other dimensional parameter is present. It would then be of great interest to obtain its large $N$ dynamics. 

For an even number of hermitiean matrices (or an arbitrary number of complex matrices) a closed sub sector\footnote{In the sense of closure under Schwinger-Dyson equations} dependent on a single matrix only, that has properties expected of such radial matrix, has indeed been identified \cite{Masuku:2009qf} \cite{Masuku:2011pm}, and it is the purpose of this communication to discuss its large $N$ dynamics.

In Section 2, the radial sector, or the restriction of the system (\ref{FIntegral}) to this closed subsector, is considered for an arbitrary number of complex matrices. The large $N$ density of radial eigenvalues is obtained explicitly and shown to have a well defined strong coupling limit satisfying (\ref{AdSOne}). A new feature present only in systems of strictly more than one complex matrix (as is relevant for $\cal{N}=$ $4$ SYM) is the appearance of a logarithmic potential that moves the eigenvalues away from the origin. In Section 3, we integrate out the angular degrees of freedom in (\ref{FIntegral}). This is carried out for a single complex matrix. In this case, the harmonic term is required, introducing another dimensionful parameter. By first mapping the system to a two matrix system, the large N limit of the radial eigenvalue distribution is obtained and shown to be of the Wigner type.    

A review of the vast literature on matrix models in beyond of the scope of this brief communication. Aspects related to the emergence of geometries are reviewed in \cite{Koch:2009gq}. 

\noindent
\section{Radial Sector}

We start by describing the closed subsector of systems of an even number of hermitian matrices which is naturally associated with a radial sector of the theory. First, we complexify by introducing complex matrices:

$$
            Z_1= X_1+iX_2 \quad , Z_2= X_3+iX_4 \quad , \text{etc} .
$$

\noindent
and in general, for $m$ complex matrices  $Z_A ~,A=1,...,m $, we consider the matrix,

\be\label{MatF}
\sum_{A=1}^m Z_A^{\dagger}Z_A ~.
\ee

\noindent
This matrix is hermitean and positive definite, and its eigenvalues

$$
\rho_i = r_i^2 ~,~ i=1,...,N ~,~ \rho_i \ge 0,
$$

\noindent
have a natural interpretation as matrix radial coordinates.

The radially invariant sector of (\ref{FIntegral}) is then:
  
$$
S_{R} = \frac{w^2}{2 g_{YM}^2} \tr \sum_{A=1}^m Z_A^{\dagger}Z_A 
+ \frac{1}{2 g_{YM}^2} \tr (\sum_{A=1}^m Z_A^{\dagger}Z_A)^2
$$

\noindent
This sector has an enhanced 
$U(N)^{m+1}$ symmetry

\be\label{sym}
              Z_A \to V_A Z_A V^{\dagger}  ~, ~ A=1,...,m,
\ee

\noindent 
and depends only on the radial eigenvalues. 

With a parametrization of the complex matrices $Z_A, A=1,..,m$ in terms of a matrix valued radial matrix and $2m-1$ unitary matrices 
\footnote{For an explicit such parametrizations in the $m=1$ case, see \cite{Masuku:2009qf}.}, one can write

$$
\int \prod_A \prod_{ij} {{dZ_A}^\dagger}_{ij} {dZ_A}_{ij} = \int \prod_i {d\rho_i} {\cal{J}}(\rho_i) d[\text{Angular}]  
$$ 
The "angular" degrees of freedom can be integrated out in the radial sector. 

${\cal{J}}(\rho_i)$ has recently been obtained in closed form \cite{Masuku:2011pm}. 
This results from the remarkable fact that correlators in this sector, with the enhanced symmetry (\ref{sym}), close under Schwinger-Dyson equations. 
The result is:

\bea
{\cal{J}}(\rho_i) &=&
C_m  \prod_i d\rho_i\rho_i^{m-1} \prod_{i > j} \rho_i^{m-1} \rho_j^{m-1} (\rho_i - \rho_j)^2 \nonumber \\
&=&
D_m  \prod_i dr_i r_i^{2m-1} \prod_{i > j} r_i^{2m-2} r_j^{2m-2} (r_i^2 - r_j^2)^2 \\
&=&  C_m \prod_i d\rho_i\rho_i^{m-1} \Delta_{RM}^2(\rho_i) = D_m \prod_i dr_i r_i^{2m-1} \Delta_{RM}^2(r_i^2) \nonumber ,
\eea

\noindent
where the antisymmetric product

$$
\Delta_{RM}(\rho_i) \equiv \prod_{i > j}  \rho_i^{\frac{m-1}{2}} \rho_j^{\frac{m-1}{2}} (\rho_i - \rho_j) 
$$

\noindent
generalizes the well known Van der Monde determinant $\Delta = \prod_{i > j} (\rho_i - \rho_j)$, and $C_m$ and $D_m$ are numerical constants.

In the radial sector, we are then lead to the integral

\be\label{radint}
Z=\int \prod_i {d\rho_i} {\cal{J}}(\rho_i)~e^{- S_R(\rho_i)} = \int \prod_i {d\rho_i} e^{\ln {\cal{J}}(\rho_i)- S_R(\rho_i)} 
= \int \prod_i~e^{- S_{eff}(\rho_i)}.
\ee

\section{Large $N$ configuration and hyperspheres}

The large $N$ configuration of (\ref{radint}) is given by the "saddle point" equation of $S_{eff}(\rho_i)$,

\be 
           S_{eff}(\rho_i) = \frac{N w^2}{2 \lambda} \sum_{i} \rho_i 
+ \frac{N}{2 \lambda} \sum_{i} \rho_i^2 - N (m-1) \sum_{i} \ln \rho_i - \sum_{i \ne j} \ln |\rho_i-\rho_j|
\ee
and satisfies:

\be\label{eqmotdisc}
2 \sum_{j, j \ne i} \frac{1}{\rho_i-\rho_j} + \frac{N(m-1)}{\rho_i} = \frac{N w^2}{2 \lambda} + \frac{N}{\lambda} \rho_i
\ee
where $\lambda=g_{YM}^2 N$ is the t'Hooft's coupling. In terms of a density of eigenvalues $\Phi(\rho)$ normalized ($\Phi \to N \Phi$) so that 
$\int d\rho \Phi(\rho)=1$, 

\be\label{eqmotcont}
\dashint _0^{\infty}\frac{d\rho' \Phi(\rho')}{\rho-\rho'} + \frac{(m-1)}{2 \rho} = \frac{w^2}{4 \lambda} + \frac{\rho}{2\lambda} 
\ee

One observes that for $m \ge 2$, and in opposition to the single hermitean and single complex matrix case, there is, in addition to the repulsive logarithmic potential amongst the eigenvalues, a new single logarithmic potential which moves the eigenvalues away from the origin. The density of radial eigenvalues therefore is non-vanishing only between hyperspheres of radii $r_-=\sqrt{\rho_-}$ and  $r_+=\sqrt{\rho_+}$, \textit{i.e.}, within a "hyperannulus". For $m=1$, and in the strong coupling limit, the density of radial eigenvalues is non-vanishing within a hypersphere.


Inspection of either (\ref{eqmotdisc}) or (\ref{eqmotcont}) establishes that the size of these hyperspheres in the strong coupling limit, \textit{i.e.}, 
when $w \to 0$, is set by the only parameter $\lambda$ left in theory through a radius $R$ satisfying

\be\label{scaling}
     R^2 \sim \rho \sim \sqrt \lambda ~, \qquad   R \sim \lambda^{\frac{1}{4}}                      
\ee


The precise form of the density and value of $\rho_{\pm}$, is obtained using standard methods \cite{Brezin:1977sv}, together with a careful treatment of the $\rho \to 0$ behaviour, following Tan \cite{Tan:1991ay}. 
The solution is completely determined by the function $G(z)$, defined on the complex plane:

$$
         G(z) = \int_{\rho_-}^{\rho_+} \frac{d\rho' \Phi(\rho')}{z-\rho'}  
              = \frac{z}{2\lambda} - \frac{(m-1)}{2z} -  \frac{q_0+\frac{z}{\lambda}}{2z} \sqrt{(z-\rho_-)(z-\rho_+)} 
$$

The absence of a pole as $z\to 0$ and the normalization of the density require:

\bea
                     & & q_0 = \frac{1}{2\lambda}(\rho_++\rho_-)~,  \quad q_0\sqrt{\rho_+\rho_-}= m-1 ~,  \nonumber \\
                     & & -\frac{m-1}{2} + \frac{q_0}{4}(\rho_++\rho_-)+\frac{1}{16\lambda} (\rho_+-\rho_-)^2 = 1  
\eea

With $s\equiv \rho_++\rho_-$ and $\Delta \equiv \rho_+-\rho_-$ these can be solved to yield

\bea\label{endpoints}
                        s^2&=&\lambda \frac{4}{3} (m+1) \left[ 1+\sqrt{1+3\left(\frac{m-1}{m+1}\right)^2} \right] \nonumber \\
                        \Delta^2 &=& s^2 - \frac{16 \lambda^2 (m-1)^2}{s^2}
\eea


Using these expressions for $s$ and $\Delta$, one has, for instance: 

\bea
\rho_{-}=0.416\lambda^{1/2},\qquad\rho_{+}&=3.10\lambda^{1/2},\qquad m=3\nonumber\\
\rho_{-}=0.168\lambda^{1/2},\qquad\rho_{+}&=2.77\lambda^{1/2},\qquad m=2\label{explicit}\\
\rho_{-}=0,\qquad\rho_{+}&=\frac{4}{3^{1/2}}\lambda^{1/2},\qquad m=1\nonumber.
\eea

In order to confirm the $\rho$ prescription around the pole, particularly for the single complex matrix case ($m=1$), while also providing a unified description for matrix systems with more than two complex matrices ($m\ge 2$), we extend the domain of definition of the density of eigenvalues to the real line. With $\rho=r^2, ~r>0$, we define
\be\label{defphir}
   2 r \Phi(r^2)  \equiv \phi(r) \equiv \phi(-r),
\ee

\noindent
so that for an arbitrary function $f(r^2)$

\be\label{asymp}
      \int_{-\infty}^{+\infty} dr f(r^2) \phi(r) = 2 \int_{0}^{+\infty} d\rho f(\rho) \Phi(\rho) .
\ee

\noindent
We remark that  ($\rho=r^2$)

\bea
\dashint_0^{\infty} \frac{d\rho' \Phi(\rho')}{\rho-\rho'} &=& \dashint_0^{\infty} \frac{2 r' dr' \Phi({r'}^2)}{r^2-{r'}^2}
 = \frac{1}{2r} \dashint_0^{\infty} dr' \phi({r'}) ( \frac{1}{r-{r'}}+ \frac{1}{r+{r'}}   ) \nonumber \\
&=&
\frac{1}{2r} \dashint_{-\infty}^{\infty} \frac{dr' \phi({r'})}{r-{r'}}
\eea

\noindent
As a result, (\ref{eqmotcont}) is equivalently written as

\be\label{eqmotr}
  \dashint_{-\infty}^{\infty} \frac{dr' \phi(r')}{r-r'} = \frac{w^2 r}{2\lambda} + \frac{r^3}{\lambda} - \frac{(m-1)}{r} ~, \qquad
\int_{-\infty}^{+\infty} dr \phi(r) = 2 
\ee

The solution of (\ref{eqmotr}) for $m>2$ is symmetric, and it can be obtained following the analysis of \cite{Tan:1991ay} \cite{deMelloKoch:1994ir}. It is a two cut solution, with the cuts in the intervals $[ -{r_+}, -{r_-}]$ and $[ {r_-}, {r_+}]$, with ${r_+}>{r_-}>0$. One finds that $\rho_{\pm}=r^2_{\pm}$, with $\rho_{\pm}$ given by equations (\ref{endpoints}) and that the densities are related by (\ref{defphir})

The solution of (\ref{eqmotr}) when $m=1$ is the standard symmetric single cut solution associated with a quartic interaction. It follows that equations (\ref{endpoints}) for the turning points indeed extend to $m=1$ provided one sets $\rho_{-}=r_{-}^2=0$. Explicitly,

$$
\pi \Phi(\rho) = \frac{1}{2 \lambda \rho} \left(\rho + \frac{1}{2} (\rho_++\rho_-)\right)\sqrt{(\rho_+-\rho)(\rho-\rho_-)} ~,~ \rho_- \le \rho \le \rho_+
$$

and

$$
\pi \phi(r) = \frac{1}{\lambda r} \left( r^2 + \frac{1}{2} (r_+^2+ r_-^2) \right)\sqrt{(r_+^2-r^2)(r^2-r_-^2)} ~,~ r_-^2 \le r^2 \le r_+^2,
$$
with $\rho_{\pm}=r_{\pm}^2$ given by (\ref{endpoints}), (\ref{explicit}).

\section{Angular degrees of freedom}

In the strong coupling limit, the scaling relationship (\ref{scaling}) is unaffected by the angular degrees of freedom, as it follows from dimensional analysis.  

For a single complex matrix ($m=1$), the integration over the angular degrees of freedom in \ref{FIntegral} can be carried out explicitly. Introducing matrix valued "polar coordinates"   

\be\label{ZRU}
  X_1 + i X_2 = Z = R U \quad , \quad Z^{\dagger} = U^{\dagger} R
\ee

\noindent
with $R$ hermitean and $U$ unitary, $R$ can be diagonalized as $R=V^{\dagger} r V$, with $r$ a diagonal matrix and $V$ unitary, we obtain \cite{Mtho}, \cite{Masuku:2009qf} 



\bea
\int \prod_A \prod_{ij} {{dZ_A}^\dagger}_{ij} {dZ_A}_{ij} &=& C_{m=1} \int \prod_i d\rho_i \Delta^2(\rho_i) [dX] [dS] \\
dX &\equiv& V dU U^{\dagger}V^{\dagger} ~,~ dS \equiv dV V^{\dagger} \nonumber 
\eea

As a result of the properties of the de Haar measure, one has  (the equality sign below is to be understood up to numerical constants), 

\bea\label{rhopartition}
Z &=& \int [dX_I] \exp\{-{\frac{w^2}{2 g_{YM}^2} \sum_{I=1}^2 \tr X_I^2 + \frac{1}{g_{YM}^2} \tr ([X_1,X_2][X_1,X_2])}\} \nonumber \\
  &=& \int [{dZ}^\dagger] [dZ] \exp\{-{\frac{N w^2}{2 \lambda} \tr Z^{\dagger}Z 
- \frac{N}{2 \lambda} (\tr ((Z^{\dagger}Z)^2) - \tr ({Z^{\dagger}}^2 Z^2))}\} \nonumber \\
 &=& \int \prod_i d\rho_i \Delta^2(\rho_i) e^{-\frac{N w^2}{2 \lambda} \sum \rho_i 
- \frac{N}{2 \lambda} \sum \rho_i^2 } \int dU e^{ \frac{N}{2 \lambda} \tr (R^2 U^{\dagger} R^2 U)} 
\label{PartRho}\\
 &=& \int \prod_i d\rho_i \Delta^2(\rho_i) e^{-\frac{N w^2}{2 \lambda} \sum \rho_i 
- \frac{N}{2 \lambda} \sum \rho_i^2 } \exp \{{N^2 X({\rho_i},{\rho_i}, \frac{1}{2 \lambda})}\}
\eea

The last integral is the well known Harish-Chandra- Itzykson-Zuber (HCIZ) integral \cite{Itzykson:1979fi}, and it can be written in closed form in terms of a determinant. We recall that, in the large $N$ limit,  $X({\rho_i},{\rho_i}, \frac{1}{2 \lambda})$ has an expansion in $1/\lambda$:

\be\label{expa}
X({\rho_i},{\rho_i}, \frac{1}{2 \lambda})= \sum_{k=1}^{\infty}\frac{1}{k} (\frac{1}{2 \lambda})^k X_k({\rho_i},{\rho_i})~,
\ee 

\noindent
where $X_k({\rho_i},{\rho_i})$ is generically a function of the weights $1/N \sum_i \rho_i^n$, and is homogeneous of degree $k$ in each of its variables. It follows that, in the strong coupling limit, should the integral be well defined, the size of the system remains determined by 

\be\
     R^2 \sim \rho \sim \sqrt \lambda ~, \qquad   R =  \lambda^{\frac{1}{4}}                      
\ee

It is of particular interest to consider the leading $1/\lambda$ contribution in {\ref{expa}}, as this is of the same order as the radially symmetric term in the action. With the leading term in the expansion (\ref{expa}) included, the effective action is now:

\be 
           S_{eff}(\rho_i) = \frac{N w^2}{2 \lambda} \sum_{i} \rho_i 
+ \frac{N}{2 \lambda} \sum_{i} \rho_i^2  - \frac{1}{2 \lambda} (\sum_{i} \rho_i)^2 - \sum_{i \ne j} \ln |\rho_i-\rho_j|
\ee

After rescaling and extending to the full real line, the saddle point equation now takes the form:

\be\label{eqmotrnew}
  \dashint_{-\infty}^{\infty} \frac{dr' \phi(r')}{r-r'} = \frac{w^2 r}{2\lambda} + \frac{r^3}{\lambda} - \frac{r w_2}{\lambda} ~,
\int_{-\infty}^{+\infty} dr \phi(r) = 2 ~, \int_{-\infty}^{+\infty} dr r^2 \phi(r) = 2 w_2
\ee

The density of radial eigenvalues is readily obtained to be:

\bea
\pi \phi(r) &=& \frac{r^2}{\lambda} \sqrt{r_+^2-r^2} ~,~ r_+= 2 {\lambda}^{\frac{1}{4}} ~, -r_+ \le r \le r_+ \nonumber \\
\pi \Phi(\rho) &=& \frac{1}{2 \lambda} \sqrt{\rho(\rho_+-\rho)} ~,~ \rho_+= 4 {\lambda}^{\frac{1}{2}} ~, 0 \le \rho \le \rho_+
\eea

\noindent
To move beyond a non-perturbative solution, we note that, as mentioned earlier, the integral can be written in closed form in terms of a determinant. Specifically,  
$$
Z=Const.\int\prod_{i}d\rho_{i}e^{-\left(\frac{N\omega^{2}}{2\lambda}\sum_{i}\rho_{i}+\frac{N}{2\lambda}\sum_{i}\rho_{i}^{2}\right)}\det\left(e^{\frac{N}{2\lambda}\rho_{i}\rho_{j}}\right).
$$

Note the cancellation of the Van der Monde determinant, i.e, the standard Coulombic-gas repulsion of the eigenvalues is absent. It turns out that this form of the partition function is not useful to obtain the large N distribution of eigenvalues. In order to facilitate this, we introduce an auxiliary two-matrix system in the next section. 

\section{Auxiliary two-matrix integral}

In $\rho$-space\footnote{In what follows, we will refer to the matrix integral expressed in
terms of the matrices $\rho$ and $U$ as the matrix integral in $\rho$-space.
}, the partition function (\ref{PartRho}) can be written as 

\begin{flalign}
Z & =\int\left[d\rho\right]\int dUe^{S_{\rho}}\\
 & =\int\prod_{i}d\rho_{i}\Delta^{2} \left(\rho\right) e^{-NV_{\rho}\left(\rho\right)}  \int dUe^{\beta N\mathrm{Tr}\rho U\rho U^{\dagger}}.
\end{flalign}
with 

\begin{equation}
S_{\rho}=-NV_{\rho}\left(\rho\right)+\beta N\mathrm{Tr}\rho U\rho U^{\dagger},\label{eq:action in rho}
\end{equation}

Motivated by the appearance of a logarithim potential for a larger number of matrices, we have in mind a potential of the form ($\epsilon >0$):

\begin{equation}
V_{\rho}\left(\rho\right)=\frac{w^{2}}{2\lambda}\mathrm{Tr}\rho+\frac{1}{2\lambda}\mathrm{Tr}\rho^{2}-\epsilon\mathrm{Tr}\ln\rho.
\end{equation}

For the saddle-point equation, we get

\begin{equation}
0=-NV_{\rho}'\left(\rho\right)+\frac{\partial}{\partial\rho_{i}}\ln\Delta^{2}\left(\rho\right)+\beta N\left\langle U\rho U^{\dagger}+U^{\dagger}\rho U\right\rangle .\label{eq:saddlepoint equation in rho space.}
\end{equation}
where $<.>$ denotes the expectation value with respect to the angular degrees of freedom.


Let us now consider the auxiliary two (hermitean)-matrix integral defined via
the action:

\begin{equation}
S=-N\mathrm{Tr}V\left(A\right)-N\mathrm{Tr}V\left(B\right)+N\mathrm{Tr}AUBU^{\dagger}.\label{eq:auxiliary}
\end{equation}
The partition function for this auxiliary two-matrix model reads

\begin{equation}\label{TwoPart}
Z=\int dA\int dB\int dUe^{-N\mathrm{Tr}V\left(A\right)-N\mathrm{Tr}V\left(B\right)+N\mathrm{Tr}AUBU^{\dagger}},
\end{equation}
where $A$ and $B$ are $N\times N$ hermitian matrices and $dU$
is the Haar measure for the $U\left(N\right)$ unitary group.

In terms of the eigenvalues of the matrices $A$ and $B$, which will
be denoted by $a_{i}$ and $b_{i}$ $\left(i=1,2,\ldots,N\right)$,
the partition function becomes

\begin{equation}
Z=\int\prod_{i}da_{i}\int\prod_{i}db_{i}\Delta^{2}\left(a\right)\Delta^{2}\left(b\right)\int dUe^{-N\sum_{i}V\left(a_{i}\right)-N\sum_{i}V\left(b_{i}\right)+\ln I\left(a,b\right)},
\end{equation}
where $\Delta^{2}\left(a\right)$ and $\Delta^{2}\left(b\right)$
are the Van der Monde determinants and $I\left(a,b\right)$ is the HCIZ integral\footnote{Here, $a=diag\left(a_{1},a_{2},\cdots,a_{N}\right)$ and $b=diag\left(b_{1},b_{2},\cdots,b_{N}\right)$.%
}:

\begin{equation}
I\left(a,b\right)=\int dUe^{N\beta\mathrm{Tr}\left(aUbU^{\dagger}\right)}.
\end{equation}

At large N, the saddle-point equations are:
\begin{flalign}
0 & =-NV'\left(a_{i}\right)+\frac{\partial}{\partial a_{i}}\ln\Delta^{2}\left(a\right)+N\left\langle UbU^{\dagger}\right\rangle \label{eq:SPE for a}\\
0 & =-NV'\left(b_{i}\right)+\frac{\partial}{\partial b_{i}}\ln\Delta^{2}\left(b\right)+N\left\langle U^{\dagger}aU\right\rangle .\label{eq:SPE for b}
\end{flalign}

Due to the exchange symmetry between $A$ and $B$ under the original integral (\ref{TwoPart}), $a_{i}=b_{i}$ at the minimum. Adding (\ref{eq:SPE for a}) and (\ref{eq:SPE for b}), one obtains

\begin{equation}
0=-NV'\left(a_{i}\right)+\frac{\partial}{\partial a_{i}}\ln\Delta^{2}\left(a\right)+\frac{N}{2}\left\langle UaU^{\dagger}+U^{\dagger}aU\right\rangle .\label{eq:SPE for aux matrix model}
\end{equation}

Comparison of  (\ref{eq:SPE for aux matrix model}) and (\eqref{eq:saddlepoint equation in rho space.})
shows that the two saddle point equations agree if

\begin{equation}
2\beta=1,\qquad V'_{\rho}\left(\rho\right)=V'\left(A\right).\label{eq:conditions}
\end{equation}
This is trivially achieved by rescaling  $\rho\rightarrow\rho\lambda^{1/2}$. The action (\ref{eq:action in rho}) becomes

\begin{equation}
S=-\frac{Nw^{2}}{2\lambda^{1/2}}\mathrm{Tr}\rho-\frac{N}{2}\mathrm{Tr}\rho^{2}+N\epsilon\mathrm{Tr}\ln\rho+\frac{N}{2}\mathrm{Tr}\rho U\rho U^{\dagger}.
\end{equation}
This action clearly satisfies the condition that $2\beta=1$ - the
other condition can be easily satisfied by setting $V'\left(A\right)=V'_{\rho}\left(\rho\right)$.
Therefore, at the level of the saddle-point equations the matrix integral
(in polar coordinates) with a Yang-Mills interaction is equivalent
to an auxiliary matrix model with an action of the form

\begin{flalign}
S= & -\frac{Nw^{2}}{2\lambda^{1/2}}\mathrm{Tr}A-\frac{N}{2}\mathrm{Tr}A^{2}+N\epsilon\mathrm{Tr}\ln A+N\mathrm{Tr}AUBU^{\dagger}\nonumber \\
 & -\frac{Nw^{2}}{2\lambda^{1/2}}\mathrm{Tr}B-\frac{N}{2}\mathrm{Tr}B^{2}+N\epsilon\mathrm{Tr}\ln B.
\end{flalign}

Methods to obtain the generating function $G\left(z\right)$, and hence the eigenvalue density, for these systems were developed in \cite{Dobroliubov:1993rw}, \cite{Makeenko:1993uk} in the context of studies of the so called "induced QCD" \cite{Migdal:1992wn}. The techniques used are based on the Schwinger-Dyson equations of the system. Of particular interest is the limit $w \to 0$ while keeping $\epsilon$ small but finite, so that $\lambda$ is the only dimensionful parameter, as considered in this communication until now. Indications are that in this limit, it may not be possible to obtain an eigenvalue density with support only on the positive real line in the presence of a pole at $z=0$ \cite{Mulokwe:2013qba}.

In the following, we therefore set $\epsilon=0$, and keep the $w$ dependent harmonic potential. This is the Hoppe integral \cite{Hoppe} \cite{Kazakov:1998ji}.

\section{Radial eigenvalues density in the Hoppe model}

The action of the auxilliary system for the Hoppe integral in matrix valued "polar coordinates" (\ref{ZRU}) is then  

\begin{flalign}
S & =-\frac{Nw^{2}}{2\lambda^{1/2}}\mathrm{Tr}A-\frac{N}{2}\mathrm{Tr}A^{2}+N\mathrm{Tr}AUBU^{\dagger}\nonumber \\
 & -\frac{Nw^{2}}{2\lambda^{1/2}}\mathrm{Tr}B-\frac{N}{2}\mathrm{Tr}B^{2}.\nonumber \\
 & =\bar{w}\mathrm{Tr}\left(A+B\right)-\frac{N}{2}\mathrm{Tr}A^{2}-\frac{N}{2}\mathrm{Tr}B^{2}+N\mathrm{Tr}AUBU^{\dagger}\nonumber \\
 & =-2N\bar{w}\mathrm{^{2}Tr}B-\frac{N}{2}\mathrm{Tr}C^{2}-N\bar{w}^{2}\mathrm{Tr}C,
\end{flalign}
where $C=A-B$ and $\bar{w}^{2}=\frac{w^{2}}{2\lambda^{1/2}}$, and where the last equality is understood under the integral. Consequently,
the partition function reads

\begin{flalign}
Z & =\int dB\int dCe^{-2N\bar{w}\mathrm{^{2}Tr}B-\frac{N}{2}\mathrm{Tr}C^{2}-N\bar{w}^{2}\mathrm{Tr}C}\nonumber \\
 & =Const.(\bar{w})\int\prod_{i}db_{i}e^{S_{eff}},
\end{flalign}
where the effective action describing the large N dynamics of the matrix $B$, and hence $\rho$, is given by

\begin{equation}
S_{eff}=\sum_{i\neq j}\ln\left|b_{i}-b_{j}\right|-2N\bar{w}^{2}\sum_{i}b_{i}.
\end{equation}
The saddle-point equations are

\begin{equation}
\sum_{j,i\neq j}\frac{1}{b_{i}-b_{j}}=2N\bar{w}^{2}.
\end{equation}
In terms of the eigenvalue density $\Phi\left(b'\right)$, the saddle-point
equations read

\begin{equation}
\dashint_{0}^{\infty}\frac{db'\Phi\left(b'\right)}{b-b'}=\bar{w}^{2}.\label{eq:posite saddlepoint equations}
\end{equation}
(Note that we have rescaled the eigenvalue density \textit{i.e.} $\Phi\left(b\right)\rightarrow N\Phi\left(b\right)$.)
Letting $b=r^{2}$ and, as explained before, extending to the full real line, the saddle-point equation can be written
as

\begin{equation}
\dashint_{-\infty}^{\infty}\frac{dr'\phi\left(r'\right)}{r-r'}=2\bar{w}^{2}r.
\end{equation}
This is the well known integral equation for the Wigner distribution which, appropriately normalized, is: 
 
\begin{equation}
\pi\phi\left(r\right)=2\bar{w}^{2}\sqrt{\frac{2}{\bar{w}^{2}}-r^{2}},\qquad 0\leq r\leq\frac{\sqrt{2}}{\bar{w}}.\label{eq:density in r}
\end{equation}

One can also solve the saddle-point equation
in (\ref{eq:posite saddlepoint equations}) directly, in which case one finds:

\begin{equation}
\pi\Phi\left(\rho\right)=\bar{w}\sqrt{\frac{2}{\rho}-\bar{w}^{2}},\qquad0\leq\rho\leq\frac{\sqrt{2}}{\bar{w}}.\label{eq:density in rho}
\end{equation}

In an approach based on commuting matrices, \cite{Berenstein:2008eg} have also obtained a radial Wigner distribution for the Hoppe model. However, given our definition of a radial coordinate, there is no approximation in our result.

\section{Summary}
Starting with the integral over systems of (an even number of) hermitean matrices interacting through a Yang-Mills potential, as is relevant to the Higgs sector of $\cal{N}=$ $4$ SYM, we identified a closed subsector that can naturally be interpreted as a radial subsector of the theory. For strictly more than two hermitean matrices, the measure introduces a logarithmic potential, a new result. The density of radial eigenvalues was obtained, and for strictly more than two hermitean matrices, was shown to have support between hyperspheres, and for two hermitean matrices, within a single hypersphere. For two hermitean matrices, the integral over the angular degrees of freedom was carried out explicitely, and in the presence of an harmonic potential, the density of radial eigenvalues was shown to be of the Wigner type, without approximation.

\section{Acknowledgements}

We would like to thank Robert de Mello Koch for interest in this project and for comments. This research is supported in part by the National Research Foundation of South Africa (Grant specific unique reference number UID 85974).

\end{document}